\documentclass[12pt,epsf]{article}
\usepackage[pctex32]{graphics}
\makeatletter

\@addtoreset{equation}{section}
\newcommand{\be}{\begin{eqnarray}}
\newcommand{\ee}{\end{eqnarray}}
\newcommand{\ba}{\begin{array}}
\newcommand{\ea}{\end{array}}
\newcommand{\nn}{\nonumber}

\makeatletter \@addtoreset{equation}{section} \makeatother

\begin{document}
\vspace{1cm}
\begin{center}
~\\~\\~\\
{\bf  \LARGE Analytic Study of  First-Order  Phase Transition in Holographic Superconductor and Superfluid}

\vspace{1cm}

                      Wung-Hong Huang\\
                       Department of Physics\\
                       National Cheng Kung University\\
                       Tainan, Taiwan\\
\end{center}
\vspace{1cm}
\begin{center}{\bf  \Large ABSTRACT } \end{center}
We use the matching method to investigate the first-order phase transition in  holographic superconductor and superfluid.   We first use the simple holographic superconductor model to describe the matching method and mention how to see the first-order phase transition. Next, we study the holographic superconductor with St\"uckelberg  term and see that the analytic results indicate the existence of first-order phase transition. Finally, we study the holographic superfluid and show that the first-order phase transition can be found for some values of parameters.  We determine the critical value analytically and compare it with the previous  numerical result.
\vspace{3cm}
\begin{flushleft}
*E-mail:  whhwung@mail.ncku.edu.tw\\
\end{flushleft}
%%%%%%%%%%%%%%%%%%%%%%%
\newpage
\section{Introduction}
The AdS/CFT correspondence is a powerful tool to deal with strongly coupled systems and is regarded as one of the most remarkable discovery in string theory [1,2,3]. In the recent years, the AdS/CFT correspondence has been applied to study strongly coupled phenomena in condensed matter physics, which was initiated in [4].  Inspired by the idea of spontaneous symmetry breaking in the presence of horizon [5], the holographic superconductors established in [6] are the remarkable example where the Gauge/Gravity duality plays an important role.  

To map a superconductor to a gravity dual, it needs to  introduce temperature by considering a black hole background [2] and a condensate through a charged scalar field [4]. To reproduce the superconductor  phase diagram, the black hole  needs to have scalar hair at low temperature, but no hair at high temperature.  Since the gauge/gravity duality traditionally requires that spacetime asymptotically approach anti-de Sitter (AdS) space at infinity, the original paper [5,6] considered a Maxwell field, $A_\mu$, and a charged complex scalar field, $\Psi$, in  3+1 black-hole AdS spacetime which is dual to a 2+1 dimensional theory.  After slightly investigation it left only two ordinary differential  equations which describe two fields.  The conditions of regularity of the fields at the horizon give two conditions for the two fields on the horizon.  Depending on the observation we can choose a condition for $\Psi$ on boundary and eventually it left only a one parameter family of solutions. Thus, after assigning the parameter, for example the chemical potential, the solution  can be solved numerically. 

Recently, two analytical approaches have been proposed to analytically calculate some properties of the holographic superconductors.  The method in [7]  is based on the matching of the field expanded  near the horizon and that expanded near the asymptotic AdS region. The authors of  [8] used the variational method for the Sturm-Liouville eigenvalue problem.  Both methods had been extensively applied to study several systems in recent [9-14]

In the present paper we will use the simple matching method to investigate the phase transition in  holographic superconductor and superfluid.  In section II we first use the holographic superconductor to describe the matching method and mention how to see the first-order phase transition.   In section III we use the matching method to study the holographic superconductor with St\"uckelberg term.  The numerical property of the model had been studied in [15] and found that it will become first-order phase transition in some cases.  The analytic study of the model with St\"uckelberg term which gives second order phase transition had been studied in [11].  In this section we focus on that with the first order transition. We get the transition temperature and compare it with the numerical result of [15].

  In section IV we use the matching method to study the holographic superfluid in [16] (see also [17]).  The model has three fields and is described by three non-linear ordinary differential equations. The high nonlinear property in the differential equations makes them be intractable analytically and the investigation of integrating the equations numerically had detailed in [16]. Some analytic properties had been studied in [9] and in this paper we will focus on the  the first order transition therein. We show that the first order transition can be found for some values of parameters.  Especially, we determine the critical value analytically and compare it with the numerical result of [16].  Last section is devoted to a short conclusion.
%%%%%%%%%%%%%%%%%%%
\section {Matching Method  in Holographic Superconductor}
The holographic superconductor studied in [6] was considering the Maxwell field $A_\mu$ minimally couple to charged scalar field $\Psi$ which is described by action
\be
S=\int d^4x \sqrt{-g}\Big( -{1\over 4} F^{\mu\nu}F_{\mu\nu}-|\partial \Psi- i A_\mu|^2  +{1\over2}|\Psi|^2 \Big)
\ee
The fields are propagating on the fixed AdS black hole background
\be
ds^2 = -f(r) dt^2 +{dr^2\over f(r)}+r^2(dx^2+dy^2),~~~f(r)=r^2\Big(1-{r^3_H\over r^3}\Big)
\ee 
in which the Hawking temperature is $T={3r_H\over 4\pi}$. After taking the static gauge we choose, $A_\mu(r)=(\phi(r),0,0,0)$ and $\Psi$ is a real function $\Psi (r)$, the associated field equations are 
\be
0&=&\phi''-{2\Psi^2\over z^2(1-z^3)}\phi\\
0&=&\Psi'' - {2+z^2\over z(1-z^3)}\Psi'+\Big({\phi^2\over r_H^2(1-z^3)^2}+{2\over z^2(1-z^3)}\Big)\Psi
\ee
in which we have changed to the variable $z\equiv {r_H\over r}$.  Thus the horizon is at $z=1$ and boundary is at $z=0$.

The regular solution near boundary can be found from above two field equations. To the second oder the results are  
\be
\phi^{(0)}(z)&=&\mu -q~z + {\cal O}(z^4)\\
\Psi^{(0)}(z)&=&0+C z^2+ {\cal O}(z^4)
\ee
in which $\mu$ is the chemical potential, $\rho\equiv q r_H$ is the charge density and condensation is defined by $\langle{\cal O}\rangle  \equiv \sqrt 2~C~r_H^2$ \footnote{Case of condensation with dimension 1 is described by $\Psi^{(0)}(z)= 0+C_1 z+...$ which can be analyzed in the same way and is neglected hereafter.} 
.  

The regular solution near horizon can be found from above two field equations. To the second oder the results are  
\be
\phi^{(H)}(z)&=&0+b (1-z)+\frac{1}{3} a^2 b (1-z)^2+....\\
\Psi^{(H)}(z)&=&a-\frac{2}{3} a (1-z)-\frac{a \left(b^2+8 r_H^2\right)}{36 r_H^2} (1-z)^2 +...
\ee
The matching method is to match the fields and their derivations which are expanding about z=1 to a finite order be equal to their expanding about z=0 to a finite order at a point $0< z_m<1$.  I.e. $\phi^{(0)}( z_m)=\phi^{(H)}( z_m), ~\phi^{(0)}( z_m)'=\phi^{(H)}( z_m)',~\Psi^{(0)}( z_m)=\Psi^{(H)}( z_m),~\Psi^{(0)}( z_m)'=\Psi^{(H)}( z_m)'$.

Consider the case which is matched at $z=1/2$ we get the following equations (this is first derived in [7]): 
\be
0&=&\frac{a^2 b}{12}+\frac{b}{2}-\mu +\frac{3 \rho }{8 \pi  T}\\
0&=&-\frac{a^2 b}{3}-b+\frac{3 \rho }{4 \pi  T}\\
0&=&\frac{1}{4} \left(-\frac{a b^2}{64 \pi ^2 T^2}-\frac{2 a}{9}\right)+\frac{2 a}{3}-\frac{9~\langle{\cal O}\rangle~}{64 \sqrt{2} \pi ^2 T^2}\\
0&=&\frac{a b^2}{64 \pi ^2 T^2}+\frac{8 a}{9}-\frac{9~\langle{\cal O}\rangle~}{16 \sqrt{2} \pi ^2 T^2}
\ee
Solving above equation we find the order parameter and  critical temperature $T_A$: 
\be
\langle{\cal O}\rangle =\frac{20}{9} \pi  T\sqrt{-\frac{32}{3} \pi ^2 T^2+\frac{3 \rho}{\sqrt{7}}},~~\Rightarrow~~
T_A={3\over4} \sqrt{\frac{\rho}{2\pi ^2\sqrt 7}}\approx  0.1038\sqrt \rho
\ee
Note that the numerically critical  value is $T_N\approx  0.118\sqrt \rho$.  Thus, the error
\be\Delta T/T_N=(T_N-T_A)/T_N=12 \%
\ee
which  was first derived in [7].  We plot above $\langle{\cal O}\rangle$  vs. temperature  in the dashing line of following diagram. For comparison, we also plot in figure 1 a solid line which is that in mean field approximation and a dot line which is the system with first-order phase transition.  
\\
\\

\hfil\scalebox{1}{\includegraphics{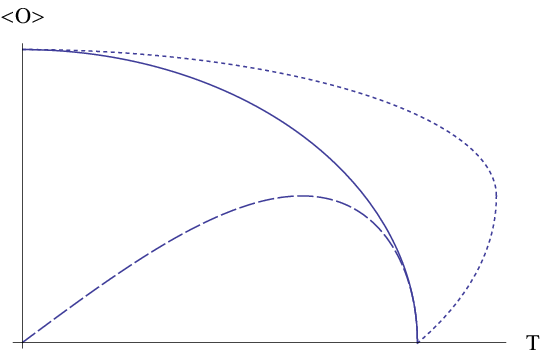}}\hfil\\
\\
{\it ~~~Fig.1. Condensation vs. Temperature.   Dashing line is the matching method result.  Solid line  is the mean field result. Dot line is the system with first-order phase transition.}
\\
\\
From above diagram we see that, while the matching method can see the university of critical exponent $1/2$ in $T_c \sim \sqrt \rho$ the value of condensation $\langle{\cal O}\rangle$ is totally wrong when $T\rightarrow 0$. 

Figure 1 tells us that there are two crucial properties for condensation $\langle{\cal O}\rangle$  in the first-order phase transition : 

(I) It have two solutions of condensation $\langle{\cal O}\rangle$ in some temperatures near $T_c$.

(II) One of  condensation $\langle{\cal O}\rangle$ is an increasing function when temperature  is near transition temperature.

These two properties will play crucial roles in following investigation about the systems with first-order phase transition.   We will see that property (I) will show in section III while property (II) will show in section IV.\\

%%%%%%%%%%%%%%%%%%%
\section {Matching Method  in Holographic Superconductor with St\"uckelberg Term}
The holographic superconductor with St\"uckelberg term studied in [15] was considering the Maxwell field $A_\mu$ minimally couple to a pair of real  scalar fields ($\Psi$, $\varphi$) which is described by action
\be
S=\int d^4x \sqrt{-g}\Big( -{1\over 4} F^{\mu\nu}F_{\mu\nu}-{1\over 2}\partial_\mu\Psi\partial^\mu\Psi +\Psi^2-{\Psi^n}|\partial \varphi- A_\mu|^2  \Big)
\ee
The fields are propagating on the fixed AdS black hole background described in section II.

After choosing gauge $\varphi=0$ and adopting $A_\mu=(\Phi,0,0,0)$ the field equations are
\be
0&=&\phi''-{2\Psi^n\over z^2(1-z^3)}\phi\\
0&=&\Psi'' - {2+z^2\over z(1-z^3)}\Psi'+{2\Psi \over z^2(1-z^3)}+{n\Psi^n~\phi^2\over r_H^2(1-z^3)^2} 
\ee
The numerical had found that above model will gives first order-phase transition  if $n>2$. In section we will discuss how to see the first-order phase transition in the analytic matching method in the case of n=3.

As before, the regular solutions near boundary and horizon can be found from above two field equations. The results are
\be
\phi^{(0)}(z)&=&\mu -\frac{3 \rho ~ z}{4 \pi  T}+....\\
\Psi^{(0)}(z)&=&\frac{9 ~\langle{\cal O}\rangle~ z^2}{16 \sqrt{2} \pi ^2 T^2}+...\\
\phi^{(H)}(z)&=&b (z-1)+\frac{1}{3} a^3 b (z-1)^2+....\\
\Psi^{(H)}(z)&=&a+\frac{2}{3} a (z-1)+ \left(-\frac{3 a^2 b^2}{64 \pi ^2 T^2}-\frac{2 a}{9}\right)(z-1)^2 +....
\ee
in which $\mu$ is the chemical potential, $\rho$ is the charge density and condensation is defined by : $\langle{\cal O}\rangle$.

We then match the above fields and their derivation at $z=1/2$ and it gives following equations: 
\be
0&=&-\mu +\frac{b}{2}+ \frac{a^3 b}{12} +\frac{3 \rho }{8 \pi  T}\\
0&=&-b-\frac{a^3 b}{3}+\frac{3 \rho }{4 \pi  T}\\
0&=&\frac{2 a}{3}+\frac{1}{4} \left(-\frac{3 a^2 b^2}{64 \pi ^2 T^2}-\frac{2 a}{9}\right)-\frac{9 ~\langle{\cal O}\rangle~}{64 \sqrt{2} \pi ^2 T^2}\\
0&=&\frac{8 a}{9}+\frac{3 a^2 b^2}{64 \pi ^2 T^2}-\frac{9 ~\langle{\cal O}\rangle~}{16 \sqrt{2} \pi ^2 T^2}
\ee
Solving above equation we find that the condensation satisfies the equation
\be
&&0=301327047 \sqrt{2} \langle{\cal O}\rangle^6+188116992000 \pi ^6 \langle{\cal O}\rangle^3 T^6\nn\\
&&~~~~~-167961600000 \pi ^6 \langle{\cal O}\rangle \rho ^2 T^6+14680064000000 \sqrt{2} \pi ^{12} T^{12}
\ee
which, however, could not be solved exactly.  After numerically solving we plot above $\langle{\cal O}\rangle$  vs. temperature  in the  following diagram. (both diagrams are very similar)
\\
\\

\hfil\scalebox{1}{\includegraphics{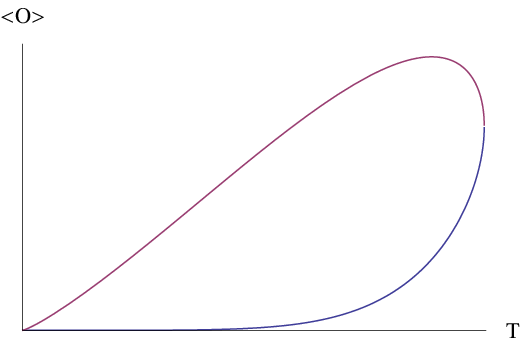}}\hfil\\
\\
{\it ~~~Fig.2. Condensation vs. Temperature in matching method.  The condensation is zero at zero temperature  is wrong.  However, the existence of two solution for a temperature indicates that it has first-order phase transition.}
\\
\\
It is clear that there appears two solutions at finite temperature, which indicates that it has first-order phase transition. Note that the condensation near T=0 becomes zero is wrong as mentioned in section II.  Nevertheless, the property that appearing double solution of condensation at a temperature in first-order phase transition still shows  in the matching method. 

To see the property from pure analysis we can consider the small value of $\langle{\cal O}\rangle$.  In this case, the first term in above equation can be neglected and it remains a three-order equation:
\be
&&0=188116992000 \pi ^6 \langle{\cal O}\rangle^3 T^6-167961600000 \pi ^6 \langle{\cal O}\rangle \rho ^2 T^6\nn\\
&&~~~~+14680064000000 \sqrt{2} \pi ^{12} T^{12}\equiv M(\langle{\cal O}\rangle)
\ee
It can be seen that $M(\langle{\cal O}\rangle)$ is positive if  $\langle{\cal O}\rangle$ =0 and it becomes negative while increasing $\langle{\cal O}\rangle$ and, eventually it becomes positive. Also, M is negative if $\langle{\cal O}\rangle$ is sufficiently negative. Above analysis thus tell us that we have two positive solutions of $\langle{\cal O}\rangle$, which indicates that the phase transition in this model is first order. As above equation is three order the solutions of condensation $\langle{\cal O}\rangle$ can be found exactly.  For example one of them is
\be
{\cal O}&=&{5 (81\times21^{1/3}\rho^2+ Z^{2/3})\over  18\times 21^{2/3}~Z^{1/3}}\\
Z&=&\sqrt{7} \sqrt{184146722816 \pi ^{12}   T^{12}-1594323 \rho ^6}-802816 \sqrt{2} \pi ^6 T^6
\ee
The analytical result tells us that only if 
\be T \le \Big({1594323\rho ^6\over 184146722816 \pi ^{12}}\Big)^{1/12}=0.121 {\sqrt \rho}\equiv T_A
\ee
can we find the double solutions of $\langle{\cal O}\rangle$. This means that the transition temperature must less then $T_A$.  The numerical  calculation in [15] found that first-order phase transition is at $T_N=0.113{\sqrt \rho}$. Thus 
\be\Delta T/T_N=(T_N-T_A)/T_N=7.1\%
\ee
Note that, in fact, we need to compare the free energy of each solution, including zero condensation. Use the property that the lowest free energy is the stable one we could then find the transition temperature.  Although the method of calculating free energy in ads/cmt was detailed in [16]  it cannot be applied in our analysis, because we can not find the analytic form of solution.

Finally, we comment that above equation tells us that solution can give $\langle{\cal O}\rangle=0$ only if T=0. This feels short of author's expectation.  Initially, we expect that we can find the solution $\langle{\cal O}\rangle=0$ near $T_c\ne 0$,  likes as that in figure 1.  Then at first-order phase transition the condensation is an increasing function near $T_C$.  While this expectation is not shown in section it is indeed shown in the holographic superfluid model studied in the next section.
%%%%%%%%%%%%%%%%%%%
\section {Matching Method  in Holographic Superfluids}
We now turn to the model of superfluid proposed and studied numerically in [16].   The model  action is the same as section II but it needs to deal with 5 fields : $A_\mu =(\phi,A_x,A_y,A_z)$ and a scalar field $\Psi$. 
\be
0&=&\phi''(z)-\frac{\Psi(z)^2 \phi(z)}{z^2 \left(1-z^3\right)}\\
0&=&A_i''(z)-\frac{3 z^2 A_i'(z)}{1-z^3}-\frac{\Psi(z)^2 A_i(z)}{\left(1-z^3\right) z^2},~~~~~A_i=A_x,A_y,A_z\\
0&=&\Psi''(z)-\frac{\left(z^3+2\right) \Psi '(z)}{z  \left(1-z^3\right)}+\frac{\phi (z)^2\Psi(z) }{r^2 \left(1-z^3\right)^2}-\frac{A_i(z)^2\Psi(z)}{r^2 \left(1-z^3\right)}+\frac{2 \Psi(z)}{z^2 \left(1-z^3\right)}
\ee
As before, the regular solutions near boundary and horizon can be found from above two field equations. The results are  
\be
\phi^{(0)}(z)&=&\mu_0 -\frac{3 \rho_0 ~ z}{4 \pi  T}+....\\
A_i^{(0)}(z)&=&\mu_i -\frac{3 \rho_i ~ z}{4 \pi  T}+....\\
\Psi^{(0)}(z)&=&\frac{9 ~\langle{\cal O}\rangle~ z^2}{16 \sqrt{2} \pi ^2 T^2} +....\\
\phi^{(H)}(z)&=&0+b0 (1-z)+\frac{1}{6} a^2 b0 (1-z)^2+....\\
A_x^{(H)}(z)&=&b1+\frac{1}{3} a^2 b1 (1-z)+\frac{a^2 b1 \left(\left(a^2+8\right) r^2+6 b1^2\right)(1-z)^2}{36 r^2}+...\\
\Psi^{(H)}(z)&=&a+a (1-z)\left(\frac{b1^2}{r^2}-\frac{2}{3}\right)\nn\\
&&+\frac{a  \left(-r^2 \left(b0^2-6 \left(a^2-2\right) b1^2\right)+9b1^4-8 r^4\right)(1-z)^2}{36 r^4}+....
\ee
in which $\mu_0$ is the chemical potential, $\rho_0$ is the charge density and condensation  is defined by $\langle{\cal O}\rangle$.   As $\mu_i$ is the  superfluid velocity and $\rho_i$ is the charge currents we will assume that the superfluid velocity and current only along x-axe.  Thus we remain only $A_x$ component.

As before, we then match the above fields and their derivations at $z=1/2$. It then gives following six equations: 
\be
0&=&-\mu_ 0+\frac{b0}{2}+\frac{3 \rho_0 }{8 \pi T}+\frac{a^2 b0}{24}\\
0&=&-b0+\frac{3 \rho_0}{4 \pi  T}-\frac{a^2 b0}{6}\\
0&=&-\mu_1+b1+\frac{a^2 b1}{6}+\frac{3 \rho_1}{8 \pi  T}+\frac{a^2 b1 \left(2 b1^2 +\frac{16}{9} \left(a^2+8\right)\pi ^2 T^2\right)}{256 \pi ^2 T^2}\\
0&=&\frac{3 \rho_1}{4\pi  T}-\frac{a^2 b1}{3}-\frac{a^2 b1 \left(2b1^2+\frac{16}{9}\left(a^2+8\right) \pi ^2 T^2\right)}{64 \pi ^2 T^2}\\
0&=&a-\frac{9 \langle{\cal O}\rangle } {64 \sqrt{2} \pi ^2 T^2}+\frac{1}{6} a \left(\frac{9b1^2}{16 \pi ^2 T^2}-2\right)\nn\\
&&+\frac{9 a \left(-\frac{16}{9} \pi ^2 T^2 \left(b0^2-2 \left(a^2-2\right)
   b1^2\right)+b1^4-\frac{2048 \pi ^4 T^4}{81}\right)}{4096 \pi ^4
   T^4}\\
0&=&-\frac{9 \langle{\cal O}\rangle}{16\sqrt{2} \pi ^2 T^2}-\frac{1}{3} a \left(\frac{9 b1^2}{16 \pi ^2 T^2}-2\right)\nn\\
&&-\frac{9 a \left(-\frac{16}{9} \pi ^2 T^2 \left(b0^2-2 \left(a^2-2\right)
  b1^2\right)+b1^4-\frac{2048 \pi ^4 T^4}{81}\right)}{1024 \pi ^4 T^4}
\ee
It is hard to analytically solve above six equations to find condensation $\langle{\cal O}\rangle$ and we will turn to find an approximation. 

 Form last two equations we see that condensation $\langle{\cal O}\rangle$ is proportional to $a$.  Therefore we will analyze the case of small condensation $\langle{\cal O}\rangle$, which may be at $T=0$ or near $T_c$ (see figure 1).   Let us detail the method in following  steps:\\

  Step 1. Using first and second equations we can delete $\rho_0$  and find solution
\be
b0 = \frac{8 \mu_0}{a^2+8}=\mu_0-\frac{a^2 \mu_0}{8}+\frac{a^4\mu_0}{64}+...
\ee

Step 2. Using third and fourth equations we can delete $\rho_1$  and find solution
\be b1 =\mu_1+ \left(-\frac{\mu_1}{2}-\frac{3\mu_1^3}{128 \pi ^2 T^2}\right)a^2 + \left(\frac{11 \mu_1}{48}+\frac{27 \mu_1^5}{16384 \pi ^4 T^4}+\frac{3
   \mu_1^3}{64 \pi ^2 T^2}\right)a^4+..
\ee
Above two results are used to express $b0$ and $b1$ in terms of $a$. 

Step 3. From fifth and sixth equations we can delete $\langle{\cal O}\rangle$.  After substituting above solutions of  $b0$ and $b1$ into it we can find $a$, which is function of $\mu_0$, $\mu_1$ and $T$. 

Step 4. Substituting this solution of $a$ into above solution of $b0$ and $b1$  we then have expressed $a$, $b0$ and $b1$ in terms of $\mu_0$, $\mu_1$ and $T$. 

Step 5.  Substituting solutions of $a$, $b0$ and $b1$ into the fifth equation we finally obtain the function of condensation:\\
\be
\langle{\cal O}\rangle&=&\frac{27 \sqrt{3} \mu_1^3 }{2048 \pi T}+\frac{\pi  T \left(3843 \mu_1^2-377 \mu_0^2\right)}{2304 \sqrt{3} \mu_1}\nn\\
&&~~~-\frac{\pi ^3 T^3 \left(943 \mu_0^4+5336 \mu_0^2 \mu_1^2+87491 \mu_1^4\right)}{5184 \sqrt{3} \mu_1^5}+O(T^4)
\ee
The above solution of condensation $\langle{\cal O}\rangle$ is not zero at T=0.   Thus, it shall be corresponding to that near $T_c$ in figure 1.  Note that, $T_c$ is the second-order phase transition point but it is not the first-order phase transition point. Also, to obtain the above relation we have expanded temperature to order $O(T^4)$, thus our result is reliable only at small T.  In fact, as T is proportional to chemical  potential $\mu_i$, our result is applicable at small of $\mu_i$. \\

Thus, if $3843\mu_1^2-377\mu_0^2 < 0$ then the condensation is a decreasing function and it corresponds to second-order phase transition (i.e. the solid line in figure 1).  On the other hand, if  $3843\mu_1^2-377\mu_0^2 > 0$ then the condensation is an increasing function and it corresponds to first-order phase transition (i.e. the dotting line in figure 1).  

Therefore we conclude that the analytically critical value of ratio $R_A$ is at
\be
R_A\equiv {\mu_1\over\mu_0}&=&\sqrt{377\over 3843}=0.3132
\ee
Above this critical value the superfluid will have first-order phase transition.  

Note that the numerical result calculated in [16] is $R_N= 0.274$.  Thus the error of ratio in our analytic analysis is 
\be
\Delta R/R_N=(R_A-R_N)/R_N=(0.3132-0.274)/0.274 \approx 14\%
\ee
Note that the error of critical temperature in analytic holographic superconductor is  $\Delta T/T_N\approx 12\%$, as mentioned in section II [7]. It is interesting to see that both errors are close. 
%%%%%%%%%%%%%%%%%%%
\section {Conclusion} 
In the present paper we use  the simple matching method [7] to investigate the phase transition in  holographic superconductor and superfluid.  We first use the holographic superconductor to describe the matching method and mention two way  to see the first-order phase transition.  The, we analytically study the holographic superconductor with St\"uckelberg term and see that there are two possible condensations at a temperature.  This indicates that the model has first-order phase transition. We get the transition temperature therein and compare it with the numerical result of [15].  We also use the matching method to study the holographic superfluids. We show that the first order transition can be found for some values of parameters and determine the critical value analytically therein.  We compare it with the previously numerical result of [16].

  The investigations of this paper show that the simple matching method [7] can be applied to study the first-order phase transition in AdS/CMT.  It is expected that the Sturm-Liouville method [8] can also be applied to investigate the first-order phase transition.  The study on the problem is under the investigation and will be presented elsewhere.
\\
\\
%%%%%%%%%%%%%%%%%%%%%%
\begin{center} {\bf REFERENCES}\end{center}
%%%%%%%%%%%%%%%%%%%%%%
\begin{enumerate}
\item  J. M. Maldacena, ``'Large N limit of superconformal field theories and supergravity," Adv. Theor. Math. Phys. 2 (1998) 231  [Int. J. Theor. Phys. 38 (1999) 1113 ] [arXiv:hep-th/9711200].
\item S. S. Gubser, I. R. Klebanov and A. M. Polyakov, ``Gauge theory theory correlators from non-critical string theory," Phys. Lett. B428 (1998) 105  [arXiv:hep-th/9802109].
\item  E. Witten, ``Anti-de Sitter space and holography," Adv. Theor. Math. Phys. 2 (1998) 253  [arXiv:hepth/9802150]. 
\item C. P. Herzog, P. Kovtun, S. Sachdev and D. T. Son, ``Quantum critical transport, duality, and Mtheory," Phys. Rev. D75 (2007) 085020  [arXiv:hep-th/0701036].
\item S. S. Gubser, ``Breaking an Abelian gauge symmetry near a black hole horizon," Phys. Rev. D78 (2008) 065034  [arXiv:0801.2977 [hep-th]]; S. S. Gubser, ``Colorful horizons with charge in anti-de Sitter space," Phys. Rev. Lett. 101 (2008) 191601  [arXiv:0803.3483 [hep-th]].
\item S. A. Hartnoll, C. P. Herzog and G. T. Horowitz, ``Holographic Superconductors," JHEP 0812 (2008) 015  [arXiv:0810.1563 [hep-th]]; G. T. Horowitz and M. M. Roberts, ``Holographic Superconductors with Various Condensates," Phys. Rev. D78 (2008) 126008  [arXiv:0810.1077 [hep-th]].
\item R. Gregory, S. Kanno and J. Soda, ``Holographic Superconductors with Higher Curva ture Corrections," JHEP 0910 (2009) 010  [arXiv:0907.3203 [hep-th]].
\item G. Siopsis and J. Therrien, ``Analytic calculation of properties of holographic superconductors," JHEP 1005 (2010) 013  [arXiv:1003.4275 [hep-th]].
\item C. P. Herzog, ``An Analytic Holographic Superconductor," Phys.Rev.D81 (2010) 126009 [arXiv:1003.3278 [hep-th]].
\item Rong-Gen Cai, Huai-Fan Li, Hai-Qing Zhang, ``Analytical Studies on Holographic Insulator/Superconductor Phase Transitions, " Phys. Rev. D83 (2011) 126007  [arXiv:1103.5568[hep-th]].
\item Chiang-Mei Chen, Ming-Fan Wu, ``An Analytic Analysis of Phase Transitions in Holographic Superconductors," Prog. Theor. Phys. 126 (2011) 387 [arXiv : 1103.5130 [hep-th]].
\item D. Momeni, Eiji Nakano, M. R. Setare, Wen-Yu Wen, ``Analytical study of critical magnetic field in a holographic superconductor, "Int. J. Mod. Phys. A 28 (2013) 1350024  [arXiv:1108.4340[hep-th]].
\item S. Gangopadhyay, D. Roychowdhury, ``Analytic study of properties of holographic p-wave superconductors," JHEP 1208 (2012) 104 [arXiv:1207.5605[hep-th]].
\item S. Gangopadhyay, D. Roychowdhury, ``Analytic study of properties of holographic superconductors in Born-Infeld electrodynamics," JHEP 1205 (2012) 002 [arXiv: 1201.6520 [hep-th]].
\item S. Franco, A. Garcia-Garcia and D. Rodriguez-Gomez,  ``A general class of holographic superconductors," JHEP 1004 (2010) 092 [arXiv:0906.1214 [hep-th]].
\item C. P. Herzog, P. K. Kovtun and D. T. Son, ``Holographic model of superfluidity," Phys. Rev. D79 (2009) 066002 [arXiv:0809.4870 [hep-th]].
\item P. Basu, A. Mukherjee, Hsien-Hang Shieh, ``Supercurrent: Vector Hair for an AdS Black Hole," Phys.Rev.D79 (2009) 045010 [arXiv:0809.4494 [hep-th]].
\end{enumerate}
\end{document}